\documentclass[12pt]{revtex4}
\usepackage{amsfonts}
\usepackage{amsmath}
\usepackage{amssymb}

\usepackage{graphicx}
\usepackage{epsfig}

\begin{document}

\title{Multielectron processes in endohedral atoms at high energies}
\author{M. Ya. Amusia$^1,^2$, E.G. Drukarev$^3$}
\affiliation{$^1$The Racah Institute of Physics,
The Hebrew University of Jerusalem, Jerusalem 91904 Israel\\
$^2$ A. F.Ioffe Physical-Technical Institute, St. Petersburg 194021 Russia\\
$^3$ B. P. Konstantinov Petersburg Nuclear Physics Institute, Gatchina,
St. Petersburg, 188300 Russia}


\begin{abstract}
We analyze the role of shake-off and of the final state interactions in  inelastic processes in the fullerene shell which follow ionization of the caged atom. We demonstrate that in the broad interval of the photon energies the process is dominated by the final state interactions. Its contribution is calculated in a model-independent way. In the large energy intervals the cross section of the process is close to that of photoionization
of isolated atom.
\end{abstract}
\maketitle

\section{Introduction}

Until now, to the best of our knowledge, the papers on the multiple  photoionization (mainly the double photoionization) of the endohedral
atoms, i.e. of the systems which consist of an atom A caged inside the fullerene shell, focused on the case when all photoelectrons where ejected
from the caged atom \cite{1}, \cite{2}. Here we consider another
channel in which ejection of one electron
from the caged atom is accompanied by inelastic multielectron processes
in the fullerene shell (FS). These may be a single-electron excitation or ionization,
double ionization, etc. We focus on the case when the
energies of the photoelectrons ejected from the internal atom $E=\omega-I_a$ are large enough
\begin{equation}
E \gg 1.
\label{1}
\end{equation}
Here $\omega$ and $I_a$ are the energies of the photon and of the
ionization of the caged atom correspondingly (we employ the atomic system of units with $e=m=\hbar=1$).
We assume that the radius of $R$ of the fullerene shell (FS) is much larger than the size of the
ionized state of the caged atom $r_{a}$,
\begin{equation}
R\gg r_{a}
\label{12}
\end{equation}%
We consider the spherical fullerens with the thickness of the shell
\begin{equation}
\Delta \ll R
\label{13}
\end{equation}%
However, the caged atom can be shifted from the center of the sphere.

This process interferes with the one in which the photon knocks out a FS electron, and the latter ionizes the internal atom.
This mechanism requires a special direction for the momentum of the electron ejected from the FS. The probability is quenched by a small factor of the order $r_a^2/R^2$. Thus we neglect the contribution of this mechanism.

As well as in the high energy double photoionization of atoms
\cite{3}, we can separate three mechanisms of the process in
endohedrals,i.e. in the systems which consist of atoms $A$ surrounded by  FS, denoted as $A@C_{N}$.
In the shake-off (SO) mechanism an atomic electron is moved to
continuum due to its direct interaction with the incoming photon,
while the second electron is ejected from the FS to continuum by the
sudden change of the effective field of the atom caged inside FS in
the endohedral. In the final state interaction (FSI) mechanism the
second electron is knocked to continuum due to direct interaction of
the ionized atomic electron with an electron belonging to FS. In the
quasifree mechanism (QFM), which was predicted long ago \cite{3}
and discovered recently in the experiments \cite{4}, absorbtion of
the photon by the two-electron system takes place almost without
participation of the nucleus.

In the double photoionization of atoms the SO mechanism dominates in the
high energy nonrelativistic asymptotics. This manifests itself in the
behavior of the double-to -single cross sections ratio $R(\omega )=\sigma
^{++}(\omega )/\sigma ^{+}(\omega )=const$. At the energies below the
nonrelativistic asymptotic both cross section $\sigma ^{++}$ and its
spectrum are the results of interplay between the SO and FSI. this is sometimes referred to as
the intermediate energy region \cite{5}.Inclusion of the QFM can be
viewed as taking into account of the lowest relativistic corrections to the
photon-electron interaction vertex \cite{6}. It leads to increase of the
ratio $R(\omega )$\cite{7}.

We shall analyze the interplay of these mechanisms in the
photoionization of the caged atom followed by excitations of the FS. Since the QFM requires the
coalescence of the two bound electrons \cite{6}, it does not contribute to the
considered channel. Thus we consider only the interplay between the SO
and the FSI.

Since the FS electrons are separated from the caged atom by the distances of the order $R \gg 1$, all
their interactions and their changes after the photoionization are quenched by a factor of the order
$1/R$, and the probability of the SO is of the order $1/R^2$.

The FSI is determined by the Sommerfeld parameter of interaction between
the fast electron moving with momentum ${\bf p}$ and those of the FS \cite{LL}
\begin{equation}
\xi=\frac{1}{pc} \ll 1,
\label{04}
\end{equation}
with $p=|{\bf p}|$, $c$ is the speed of light. Since the binding energies of the valence FS electrons $I_{FS} \leq 1$, we can write
\begin{equation}
\xi^2 \approx \frac{1}{2E} \ll 1.
\label{04a}
\end{equation}

We include the FSI terms of the order $\xi^2$ employing the technique worked out in \cite{7a}.
Note, however that the probability of FSI contains also the number of the active FS electrons $N$, i.e.
the real parameter is $\xi^2N$. In the SO the FS reacts on the change of the field as a whole and the probability does not contain the factor $N$.

Because of lack of information about the FS wave function we calculate only the sum of cross sections of
inelastic processes called also the cross section of absorption, in which ionization of the internal atom is followed by ionization of the FS or its transitions to  excited states, double ionization of FS, etc.

The cross section of the process with transition of the FS to a particular final state $n$
contains the cross section of photoionization of the isolated atom $\sigma_{\gamma}$ as a factor.
The factorization is violated by the terms of the order $V/E$ with $V$ the potential of the FS "felt"
by the photoelectron. Since $|V| \leq 1 Ry$, such terms can be neglected in our approach.
\begin{equation}
\sigma_{n}=\sigma_{\gamma}S_n,
\label{05}
\end{equation}

We investigate the behavior of the ratio
\begin{equation}
r(E)=\frac{\sigma_A(E)}{\sigma_{\gamma}(E)},
\label{1a}
\end{equation}
$\sigma_{\gamma}$ is the cross section of photoionization of the isolated atom in which the photoelectron
carries the energy $E=\omega-I_a$. We calculate the absorption cross section $\sigma_A$ as the difference between the total cross section
\begin{equation}
\sigma_t=\sigma_{\gamma}S_t ; \quad S_t=\sum_nS_n
\label{1aa}
\end{equation}
and the elastic cross section  $\sigma_0=\sigma_{\gamma}S_0$, i.e.
\begin{equation}
r(E)=S_t(E)-S_0(E)
\label{1b}
\end{equation}

For the SO mechanism this requires the knowledge of the ground state wave functions of the FS. We suggest a simple model which assumes that the FS electrons have a uniform distribution
inside the shell. The contribution of the FSI can be calculated in a model-independent way. It depends only on the number of the active electrons in the FS. We calculate the FSI beyond the perturbative approach \cite {70} and trace the energy dependence
of the function $r(E)$ determined by Eq.(\ref{1b}).

\section{Shake off}

\subsection{General equation}

We can write for the amplitude of the double photoionization in the
considered channel
\begin{equation}
F_{SO}=F_{\gamma }\langle{\Phi}_{x}|\Psi _{0}\rangle .
\label{21}
\end{equation}
Here $F_{\gamma }$ is the amplitude of photoionization of $n$-th atomic
state, $\Psi _{0}$ describes ground state of the FS electrons moving in the superposition of
its self-consistent field and that of the internal atom. In the final state
$\Phi_{x}$ one of the FS electrons is moved to the continuum.
the electrons feel the self-consistent field of the FS and that of
the ion with the hole in the $n$-th state of its electronic shell.
Here one of the FS electrons is moved to the continuum.
The matrix element on the RHS of Eq.(\ref{21}) obtains nonzero values only if
the angular momenta the initial and final states have the same angular momenta.
Thus the SO can lead only to the monopole transitions.

The sum of the cross sections of inelastic processes can be written as
\begin{equation}
\sigma_{SO}=\sigma_{\gamma }(1-\langle{\Phi}_{0}|\Psi _{0}\rangle^2),
\label{22}
\end{equation}
with $\sigma_{\gamma}$ the cross section of the photoionization,
$|\Phi _{0}\rangle$ is the ground state of the FS with a hole in the electronic shell of the caged atom created by the photon impact.

\subsection{A model for the ground state}

The wave function $\Psi _{0}$ is strongly quenched outside the
region
\begin{equation}
R\leq r\leq R+\Delta,
\label{23}
\end{equation}
with the radial part depending on
\begin{equation}
x=r-R;\quad 0\leq x\leq \Delta .
\label{24}
\end{equation}
The same refers to the function $\Phi_0$. However the values of the parameters $R$ and $\Delta$ in the state
$|\Psi _{0}\rangle$ differ from those in the state $|\Psi _{0}\rangle$.

To estimate the matrix element  $\langle{\Phi}_{0}|\Psi _{0}\rangle$ in Eq.(\ref{22}) we assume that in the ground state
the FS density does not depend on $x$. Under this assumptions the wave function of the FS electrons with the angular momentum $\ell$ is
\begin{equation}
\Psi_i({\bf r})=(\frac{N_{\ell}}{V})^{1/2}Y_{\ell,m}(\Omega)
\label{25}
\end{equation}
for $r$ in the interval determined by Eq.(\ref{24}), vanishing otherwise ($\Omega$ is the solid angle).
In this equation $V=4\pi R^2\Delta$ is the volume of the FS.

The FS electrons can be viewed as moving in an effective self consistent field $U_i$ which also does not depend on $x$. The Thomas-Fermi equation
\begin{equation}
\rho_0=(2U_0)^{3/2}\frac{1}{3\pi^2},
\label{26}
\end{equation}
relates the electron density $\rho_0$ and the potential
\begin{equation}
U_0=\frac{1}{2}(3\pi^2\frac{N}{V})^{2/3}.
\label{210}
\end{equation}
Here $N$ is the total number of active electrons.

After ejection of the photoelectron a new value of the potential is
\begin{equation}
U_f=U_0+U_h,
\label{27}
\end{equation}
where $U_h$ is the potential created by the hole in the state $n$ of the internal atom
\begin{equation}
U_h(r)=\int d^{3}r_{a}\frac{\rho _{n}(\mathbf{r_{a}})}{|\mathbf{r}-\mathbf{%
r_{a}}|}\approx \frac{1}{r} \approx \frac{1}{R}.
\label{27a}
\end{equation}
where $\rho_n$ is the electron density in the state $n$,
$\mathbf{r_{a}}$ and $\mathbf{r}$ are the coordinates of the atomic
electron and of that in the FS.

Thus the final state wave function
\begin{equation}
\Phi_0({\bf r})=\sum _{\ell}\frac{N_{\ell}}{V'}Y_{\ell,m}(\Omega)
\label{28}
\end{equation}
with
\begin{equation}
V'=V+\frac{3}{2}\frac{1}{RU_0}.
\label{29}
\end{equation}
Employing Eqs.(\ref{25}) and (\ref{28}) we obtain
\begin{equation}
\langle\Phi_{0}|\Psi _{0}\rangle =1-\frac{3}{4}\frac{1}{RU_0}
\label{211}
\end{equation}

Assuming, following \cite{7b} the values of the FS parameters for the fullerene $C_{60}$
\begin{equation}
R=6.02; \quad \Delta=1.25
\label{212}
\end{equation}
(in this case $N=240$ )
we find
\begin{equation}
1-\langle\Phi_{0}|\Psi _{0}\rangle =0.046 ; \quad \langle\Phi_{0}|\Psi _{0}\rangle^2=0.91
\label{212}
\end{equation}
The numerical results do not depend strongly on the actual values of the parameters.
For example taking $R=5.75a.u.$, $\Delta =1.89a.u.$
\cite{8} we find
$$1-\langle\Phi_{0}|\Psi _{0}\rangle =0.060 ; \quad \langle\Phi_{0}|\Psi _{0}\rangle^2=0.88.$$

\section{Final state interactions in perturbative approach}

\subsection{Lowest order terms}

The amplitude of a process which includes the final state interaction between the fast electron and the electronic shell up to the terms of the order $\xi^2$ is \cite{7a}
\begin{equation}
F_x=F_x^{(0)}+F_x^{(1)}+F_x^{(2)},
\label{213}
\end{equation}
where the upper index denotes the number of interactions between the fast electron and the FS,
index $x$ labels the final state of the FS.
The amplitudes $F_x^{(i)}$ contain the amplitude of ionization of the isolated atom $F_{\gamma}$
as a factor \cite{7a}
\begin{equation}
F_{x}^{(i)}
=F_{\gamma}T^{(i)}; \quad i=0,1,2.
\label{214}
\end{equation}
Here $T^{(0)}=\langle{\Phi}_{n}|\Psi _{i}\rangle$ is the SO matrix element-see Eq.(\ref{21}).
The accuracy of this equation is $V_a/E$ with $V_a$ the potential energy of the photoelectron in the field of the caged atom.

One can write for the cross section of the process with transition of the FS to a particular final state $x$
\cite{7a}(see Eq.(\ref{05}))
\begin{equation}
S_x=|T_x^{(0)}|^2+2T_x^{(0)}ReT_x^{(1)}+|Im T_x^{(1)}|^2+2T_x^{(0)}ReT_x^{(2)},
\label{216}
\end{equation}

Now we calculate the FSI amplitudes. One can write
\begin{equation}
T_x^{(1)}=\langle\Phi_x|U_1|\Psi_0\rangle,
\label{218}
\end{equation}
where $U_1=\sum_kU_1(r^{(k)})$, with $k$ labeling the FS electron, $U_1(r^{(k)})$ is its interaction with the photoelectron in the lowest order of the FSI. One can write
\begin{equation}
U_1(\mathbf{r}^{(k)})=\frac{1}{c}\int \frac{d^{3}f}{(2\pi )^{3}}G({\bf f})g(f)
e^{i(\mathbf{f}\cdot \mathbf{r}^{(k)})},
\label{219}
\end{equation}
where $G(f)=2(\mu^2-(\mathbf{p+f)}^{2}+i\nu)$ with $\mu ^{2}=p^{2}+2\varepsilon
_{fi}$ is the free electron propagator, energy $\varepsilon_{fi}$ is transferred by the FSI, $g(f) =4\pi/(f^2+\lambda^2)$, $\lambda \rightarrow 0$. Keeping only the term proportional to the large momentum $p$ in denominator of the electron propagator we can put
\begin{equation}
G({\bf f})=\frac{-2}{2({\bf p} \cdot {\bf f})-i\nu},
\label{220}
\end{equation}

Using the well known formula
\begin{equation}
\frac{1}{a}\cdot\frac{1}{c}=\int_{0}^1\frac{dx}{\Big(ax+c(1-x)\Big)^2}
\label{222}
\end{equation}
we obtain
\begin{equation}
 -\frac{1}{2({\bf p}\cdot{\bf f})-i\nu}\cdot\frac{1}{f^2+\lambda^2}=
 -\int_{0}^1\frac{dx}{\Big(2({\bf p}\cdot{\bf f })(1-x)+f^2x+\lambda^2x-i\nu)^2}.
\label{224}
\end{equation}
Introducing $y=(1-x)/x$, ${\bf f}'={\bf f}+{\bf p}y$ and integrating over ${\bf f}'$ by using the relation
\begin{equation}
\int\frac{d^3f}{(2\pi)^3}\cdot\frac{4\pi e^{i({\bf f} \cdot{\bf r})}}{(f^2-b^2-i\nu)^2}=\frac{1}{2b}\frac{\partial}{\partial b}\frac{e^{ibr}}{r}=i\frac{e^{ibr}}{2b};\quad b^2=p^2y^2-\lambda^2
\label{225}
\end{equation}
we find
\begin{equation}
U_1({\bf r}^{(k)})=\frac{-i}{c}\int_0^{\infty} \frac{dy}{b(y)}e^{ib(y)r^{(k)}-i({\bf p} \cdot{\bf r}^{(k)})y}.
\label{226}
\end{equation}
this leads to
\begin{equation}
U_1=i\xi\sum_k\ln(r^{(k)}-r^{(k)}_z)\lambda
\label{6}
\end{equation}
Thus the amplitude $T^{(1)}$ is mostly imaginary
\begin{equation}
T_x^{(1)}=i\xi\langle\Phi_x|\sum_k\ln(r^{(k)}-r^{(k)}_z)\lambda|\Psi_0\rangle.
\label{227}
\end{equation}
The imaginary part dominates since the pole of the electron propagator provides the leading contribution.
This means that the photoelectron passes the distances  of the order of the FS radius $R \gg 1$, and interacts with the FS electrons at the region of their location.
The divergence at $\lambda=0$ is just the Coulomb phase of the interaction between the photoelectron and the electronic shell \cite{7a}. The divergent contributions will cancel after the second order terms will be taken into account.

Since the leading contribution to $T_x^{(1)}$ is imaginary, while $T_x^{(0)}$ is real,
the
leading nonvanishing contribution of the FSI is of the order $\xi^2$. In order to find it, one has to include  the second order amplitude $T_x^{(2)}$ and the terms of the order $f/p$ in the first order amplitude $T_x^{(1)}$.

The second order amplitude can be written as
\begin{equation}
T_x^{(2)}=\langle\Phi_x|U_2|\Psi_0\rangle,
\label{399}
\end{equation}
with
\begin{equation}
U_2=\frac{1}{c^2}\sum_{k_1k_2}\int \frac{d^{3}f_1}{(2\pi )^{3}}\frac{d^{3}f_2}{(2\pi )^{3}}G({\bf f_1})g(f_1)G({\bf f_1+\bf f_2})g(f_2)
e^{i(\mathbf{f}_1\cdot \mathbf{r}^{(k_1)})}e^{i(\mathbf{f}_2\cdot \mathbf{r}^{(k_2)})},
\label{400}
\end{equation}
Using Eq.(\ref{220}) for the Green function $G$ and putting in the integrand
\begin{equation}
\frac{1}{({\bf p} \cdot {\bf f}_1)}\frac{1}{({\bf p} \cdot( {\bf f}_1+{\bf f}_2))}=\frac{1}{2}\Big(\frac{1}{({\bf p} \cdot {\bf f}_1)}\frac{1}{({\bf p} \cdot( {\bf f}_1+{\bf f}_2))}+\frac{1}{({\bf p} \cdot {\bf f}_2)}\frac{1}{({\bf p} \cdot( {\bf f}_1+{\bf f}_2))}\Big)=
\label{401}
\end{equation}
$$\frac{1}{2}\frac{1}{({\bf p} \cdot {\bf f}_1)}\frac{1}{({\bf p} \cdot{\bf f}_2)}.$$
we find that
\begin{equation}
U_2=U_1^2/2
\label{402}
\end{equation}
Here again the FSI takes place in the region of the FS.
 As to the terms of the order $f/p$ in the amplitude $T_x^{(1)}$, they determine its the real
part $Re T_x^{(1)}$, they describe the interactions between the photoelectron while it is close to the caged atom and the FS. Thus they have additional factor $1/R \ll 1$. Thus the second term on the RHS of Eq.(\ref{216}) is
\begin{equation}
T_x^{(0)}ReT_x^{(1)} \approx \frac{\xi^2N}{R}|\langle\Phi_x|\Psi_0\rangle|^2 \approx \frac{\xi^2N}{R^3}
\label{406}
\end{equation}

Hence,neglecting the terms of the order $1/R$ in the FSI terms, we find $S_t=1$ while
\begin{equation}
S_0=|T_0^{(0)}|^2+|\langle\Psi_0|U_1|\Psi_0\rangle|^2-\langle\Psi_0|U_1^2|\Psi_0\rangle.
\label{404}
\end{equation}
We can put $r^{(k)}=R$ in Eq.(\ref{6}). As expected, the terms containing $\lambda$ cancel on the RHS of Eq.(\ref{404}). Hence we can employ
\begin{equation}
U_1=i\xi\Lambda; \quad {\Lambda}=\sum_k\ln(1-t^{(k)})
\label{406}
\end{equation}
with $t^{(k)}=r^{(k)}_z/r^{(k)}$.
Thus Eq.(\ref{404}) takes the form
\begin{equation}
S_0=|T_0^{(0)}|^2+\xi^2|\Big((\langle\Psi_0|\Lambda|\Psi_0\rangle|^2-\langle\Psi_0|\Lambda^2|\Psi_0\rangle\Big).
\label{406a}
\end{equation}
Here the three terms on the RHS describe the SO, FSI and their interference correspondingly.
Direct calculation provides
\begin{equation}
r(E)=\zeta+\xi^2N; \quad \zeta=1-|T_0^{(0)}|^2
\label{408}
\end{equation}

Employing of closure requires that the energy $E$ is large enough to include all important excited states
\begin{equation}
E \gg \bar{\varepsilon}.
\label{408a}
\end{equation}
At large energies $\varepsilon \gg I_{FS}$ the energy distributions drop as $1/\varepsilon^2$ and thus the energy losses $\bar{\varepsilon}$ are determined by $I_{FS} \ll \varepsilon \ll E$. They are \cite{8a}
\begin{equation}
\bar{\varepsilon}=\frac {\xi^2N}{4R^2}\ln\frac{E}{I_{FS}}
\label{408b}
\end{equation}

Some additional data can be obtained by studying the distributions of the electrons, ejected from the FS.

\subsection{Partial wave analysis}

The ratio $r(E)$ given by Eq.(\ref{408})can be written as the sum of the contributions of the partial waves
\begin{equation}
r(E)=\sum_{n\ell}|\langle\Phi_{n\ell}|\Psi_0\rangle|^2+\xi^2\sum_{n\ell}|\langle\Phi_{n\ell}|\Lambda |\Psi_0\rangle|^2-\xi^2\sum_{n\ell}\langle\Psi_0|\Phi_{n\ell}\rangle\langle\Phi_{n\ell}|\Lambda^2|\Psi_0\rangle,
\label{406b}
\end{equation}
with the sum carried out over the excited states $n\neq 0$.

The first and the third terms term on the RHS obtain  nonzero values only for $\ell=0$.
Expanding in terms of the Legandre polynomials
\begin{equation}
\ln(1-t)=\sum_{\ell}a_{\ell}P_{\ell}(t); \quad a_{\ell}=\frac{2\ell+1}{2}\int_{-1}^{1}dtP_{\ell}(t)ln(1-t),
\label{409}
\end{equation}
we see that the contributions of the terms with $\ell=0$ to the third and second terms cancel.
Thus
\begin{equation}
r(E)=\sum_n(\delta_{\ell 0}+\xi^2N\sum_{\ell=1}b_{\ell}A_{n\ell})^2,
\label{419}
\end{equation}
with
\begin{equation}
A_{n\ell}=\langle\Phi^r_{n\ell}|\Psi^r_0\rangle,
\label{420}
\end{equation}
while
\begin{equation}
b_{\ell}=\frac{2\ell+1}{\ell^2(\ell+1)^2}.
\label{308a}
\end{equation}
Or,employing closure
 \begin{equation}
r(E)=1-A^2_{00}+\xi^2N\sum_{\ell=1}b_{\ell}.
\label{419a}
\end{equation}
Presenting
\begin{equation}
\frac{2\ell+1}{\ell^2(\ell+1)^2}=\frac{1}{\ell^2}-\frac{1}{(\ell+1)^2}
\label{308b}
\end{equation}
we see that indeed $\sum b_{\ell}=1$.

The dipole term which dominates in the FSI
provides $3/4$ of its contribution.
The amplitude of the dipole transition to a particular excited state $n$ is proportional to the overlap of the radial wave functions
$ \langle\Phi_{n1}^r|\Psi^r_0\rangle$.
Thus investigation of the spectrum of the electrons, ejected from the FS would provide
the data which is complimentary to that obtained from the studies of direct
photoionization of the fullerene $C_{60}$ \cite{11},\cite{12}.

Note, however that
the  perturbative results, presented in the last  two Subsections are true if  $\xi^2N \ll 1$. In the case of the fullerene $C_{60}$ this means that
$E \gg 5 $ keV. At such energies all $N=360$ electrons of the FS can participate on the process. In order to analyze the lower energies we must go beyond the perturbative approach.

\section{Final state interactions beyond the perturbative approach}.

One can see that Eq.(\ref{401}) can be generalized for the case of arbitrary number $n$ of interactions between the photoelectron and the FS. Introducing $a_n= ({\bf p} \cdot( {\bf f}_1+{\bf f}_2+..{\bf f}_n))$ we can write
\begin{equation}
\frac{1}{a_1}\cdot \frac{1}{a_2}...\cdot\frac{1}{a_n}=\frac{1}{n!}\frac{1}{a_1^n}.
\label{409}
\end{equation}
This equation which can be proved by the induction method, was used for calculation of the radiative corrections in the electromagnetic interactions \cite{14}. Thus the amplitude of transition to the state $\langle\Phi_x|$ is
\begin{equation}
T_x=\langle\Phi_x|e^{i\xi\Lambda}|\Psi_0\rangle,
\label{410}
\end{equation}
with $\Lambda$ defined by Eq.(\ref{406}). Thus
\begin{equation}
r(E)=1-|\langle\Phi_0|e^{i\xi\Lambda }|\Psi_0\rangle|^2\approx 1-|\langle\Psi_0|e^{i\xi\Lambda}|\Psi_0\rangle|^2
\label{411}
\end{equation}
The last equality is due to the fact that the change of the field of the caged atom provides the perturbation of the
order $1/R$. Hence
\begin{equation}
r(E)=1-|\langle\Phi_0|\Pi_k(1-t^{(k)})^{i\xi}|\Psi_0\rangle|^2.
\label{412}
\end{equation}
This provides
\begin{equation}
r(E)=1-\frac{1}{(1+\xi^2)^N}|\langle\Phi_0|\Psi_0\rangle|^2=1-e^{-N\ln(1+\xi^2)}|\langle\Phi_0|\Psi_0\rangle|^2.
\label{413}
\end{equation}
Thus we can write
\begin{equation}
r(E)=\zeta+r_F(E); \quad  r_F(E)=1-e^{-N\ln(1+\xi^2)}.
\label{414}
\end{equation}
Here $\zeta$ defined by Eq.(\ref{408}) is the SO contribution which does not depend on $E$,
$r_F(E)$ is the contribution of FSI.

If the photon energy is so large that $N\xi^4 \ll 1$ ( i.e. $E \gg 300$ eV) we find
\begin{equation}
r_F(E)=1-e^{-N\xi^2}.
\label{415}
\end{equation}
At these energies $N$ is the total number of the FS electrons, i.e. $N=360$.

\section{Total cross section}

Now we can trace the energy dependence of the cross section of absorption by the FS in this process.
Employing Eqs.(\ref{408a}) and (\ref{408b}) we find that closure can be used at $E \geq 50$ eV.
If the energy $E$ is smaller than the ionization potential of the core $1s$ electrons $I_c \approx 315$ eV., the ratio
$r_F(E)$ is determined by Eq.(\ref{414}) with $N=N_v=240$ the number of the valence electrons. At $E$ close to $I_c$
we find $1-r_F(E) \approx 2\cdot 10^{-5}$.

At $E>I_{c}$ the core electrons are involved into the process as well. While
$E$ is of the order $I_s$ their contribution can not be calculated by employing closure since some of the exited states can not be reached due to the energy conservation low. However, Eq.({\ref{414}) with $N=N_v=240$ provides the lower limit for the value of $r_F(E)$ at these energies.

At larger energies $E \gg I_c$ the ratio $r_F(E)$ is determined by Eq.(\ref{415}) with the core electrons included, i.e. $N=360$. At $E=2$ keV we find $1-r_F(E) \approx 0.09$, i.e. $r_F(E)$ is still very close to unity. At $E=5$ keV
$r_F(E) \approx 0.62$, dropping as $1/E$ at larger energies, following Eq.(\ref{408}).
The FSI and the SO contributions to the ratio $r(E)$ Eq.(\ref{414}) become of the same order at $E \geq 50$ keV.

If the binding energy of the ionized state of the internal atom $I_a$ and the photon energy are small enough,
both SO and FSI contributions to the cross section are enhanced by the same factor. This happens because
external photon is strongly influenced by the FS due to polarization of
the latter. This effect manifests itself in a factor
\begin{equation}
D(\omega )=1-\alpha (\omega )/R^{3}
\label{301}
\end{equation}
in the amplitude of ionization of internal atom  \cite{9}. Here $\alpha (\omega )$ is the dynamic dipole
polarizability of the FS. Since $\alpha (\omega )<0$ the factor $|D(\omega )|^{2}$
increases the cross section of the single photoionization \cite{9}
\begin{equation}
\sigma _{pol}^{+}(\omega )=\sigma ^{+}(\omega )D^2(\omega ).
\label{302}
\end{equation}

The polarizability $\alpha (\omega )$ has a strong maximum due to FS giant
resonance at $\omega \approx 1 a.u$. \cite{9}. The role of polarization
diminishes with increasing of the photon energy.
It becomes negligible
at $\omega _{\max }\approx 2.5 a.u$,
thus here $D^2(\omega )\approx 1$.
However, the characteristic binding energy of a FS electron is $I =7 $eV. If the binding
energy of the ionized state of the caged atom is of the same order or smaller,
the factor $D^2(\omega)$ increases both SO and FSI contributions to the double photoionization cross section at the lower
limit. Note that the factor $D^2(\omega)$ enters the cross sections of the double and single photoionization in the same way. Thus it cancels in their ratio $r(E)$-Eq.(\ref{1a}).

\section{Summary}

We investigated the high energy photoionization of the
endohedral atom $A@C_{60}$ followed by inelastic processes in the fullerene shell. We
traced the energy dependence of the ratio $r(E)$ of the cross section to that of
photoionization of isolated  atom.
In a broad
interval of the values of the photoelectron energies $E$ the ratio is
dominated by the final state interactions and is calculated in a model independent way
beyond the perturbative approach. The ratio is shown to be close to unity
until we reach the region of $E$ of several keV. At the energies higher than $E=5 keV$
where $r(E)=0.62$ it drops as $1/E$. The contributions of the FSI and of the shake-off become
of the same order at very high energies $E >50$ keV.

The work was supported by the MSTI-RFBR grant 11-02-92484. One of us (EGD)
thanks for
hospitality during the visit to the Hebrew University of Jerusalem.

\end{document}